# Adaptive Inference on General Graphical Models

**Umut A. Acar**[*]
Toyota Tech. Inst.
Chicago, IL
*umut@tti-c.org*

**Alexander T. Ihler**
U.C. Irvine
Irvine, CA
*ihler@ics.uci.edu*

**Ramgopal R. Mettu**[†]
Univ. of Massachusetts
Amherst, MA
*mettu@ecs.umass.edu*

**Özgür Sümer**
Univ. of Chicago
Chicago, IL
*osumer@cs.uchicago.edu*

## Abstract

Many algorithms and applications involve repeatedly solving variations of the same inference problem; for example we may want to introduce new evidence to the model or perform updates to conditional dependencies. The goal of *adaptive inference* is to take advantage of what is preserved in the model and perform inference more rapidly than from scratch. In this paper, we describe techniques for adaptive inference on general graphs that support marginal computation and updates to the conditional probabilities and dependencies in logarithmic time. We give experimental results for an implementation of our algorithm, and demonstrate its potential performance benefit in the study of protein structure.

## 1 Introduction

It is common in many applications to repeatedly perform inference on a variations of essentially the same graphical model. For example, in a number of learning problems we may use observed data to modify a portion of the model (e.g., fitting an observed marginal distribution), and then recompute various moments of the new model before updating the model further [8]. Another example is in the study of protein structures, where a graphical model can be used to represent the conformation space of a protein structure [15, 9]. The maximum-likelihood configuration in this model then corresponds to the minimum-energy conformation for the corresponding protein. An application of interest in this setting is to perform amino acid mutations in the protein to determine the effect of these mutations to the structure and the function of the protein.

The changes described in the examples above can, of course, be handled by incorporating them into the model and then performing inference from scratch. However, in general we may wish to assess thousands of potential changes to the model; for example, the number of possible mutations in a protein structure grows exponentially with the number of considered sites. *Adaptive inference* refers to the problem of handling changes to the model (e.g. to conditional dependencies and even graph structure) more efficiently than performing inference from scratch. Delcher *et al.* [6] studied this problem under a set of fairly restrictive conditions, requiring that the graph be tree-structured and supporting only changes to the observed evidence in the model. They show that updates to observed evidence may be performed in expected $O(\log n)$ time, where $n$ is the size of the graph. More recently, Acar *et al.* [2] gave a method of supporting more general changes to the model so long as the model remains tree-structured.

Unfortunately, many graphical models of interest are not trees, but are "loopy". In principle, we can perform adaptive inference on loopy graphs by constructing their junction tree [13] and applying existing frameworks to the junction tree itself [6, 2]. This approach, however, can be very slow since even a small change to the graph can cause the junction tree to change dramatically, e.g., creating a cycle by inserting a new edge can require a linear number of changes to the junction tree.

In this paper, we present techniques for supporting adaptive inference on general graphical models efficiently. Given a factor graph $G$ with $n$ nodes, maximum degree $k$, and domain size $d$ (variables can take $d$ different values), we require the user to specify a spanning tree $T$ of $G$. We then construct a *(hierarchical) clustering* of $G$ with respect to the spanning tree $T$ (Sec. 3). The hierarchical clustering is a tree of clusters where each cluster represents a subgraph of $G$. A key property of the clustering is that it has expected $O(\log n)$ depth, where the expectations are taken over internal randomization. For each cluster we compute a *cluster function*, a partial marginalization of factors in the cluster. We show that the cluster functions can be computed in $O(\alpha^m)$ where $\alpha = d^{k+1}$ and $m$ is the size of the boundary of the cluster.

---
[*]U. A. Acar is supported by a gift from Intel.

[†]R. R. Mettu is supported by a National Science Foundation CAREER Award (IIS-0643768).

Given such a hierarchical clustering, we show how to compute the marginal at any variable by performing a traversal from the top level cluster to the variable. Since maximum path length in the clustering is expected $O(\log n)$, we show that marginals can be computed in expected $O(\alpha^\beta \log n)$ time where $\beta$ is an upper bound on the boundary size of all clusters (for a tree-structured factor graph $\beta = 2$). The novel contribution of our approach is that our clustering also allows efficient updates to factors and edge insertions/deletions in the input graph. We show that after any of these updates is applied, it is possible to update the clustering $O(\alpha^\beta \log n)$ time and that marginals computed thereafter correctly reflect the updates.

Our results generalize the previous techniques for adaptive inference with tree-structured factor graph to loopy graphs. The main insight is to partition the loopy graph into a spanning tree and a set of non-tree edges and cluster the graph based on the spanning tree only. This enables updating the hierarchical clustering in expected logarithmic time when an edge is inserted or deleted using RC-Trees. When computing marginals, contributions of the nodes of the graph are computed in the order specified by the clustering on the spanning-tree edges. Compared to previous work on factor trees [2], we also simplify marginal computations.

We note that our bounds depend exponentially on the boundary size of the clusters. While this exponential cost can be large in general, for many interesting classes of graphs it can be kept small. Moreover, since our expected running times are logarithmic in $n$, our approach can still be significantly faster than computing from scratch. This exponential factor is not surprising, since exact inference on general graphs is NP-hard; conventional, algorithms for exact inference also have an exponential dependence on some property of the graph such as the tree width.

To evaluate the effectiveness of the proposed techniques, we implemented our algorithm and compared its performance against an implementation of sum-product that performs inference on a junction-tree of the given factor graph. Our experiments on a synthetic benchmark for factor graphs show that our approach can be orders of magnitude faster than sum-product. We also investigate the applicability of our algorithm to study protein structure, and show that our algorithm is considerable faster than sum-product for modeling several moderately-sized proteins.

## 2 Background

Graphical models provide a convenient formalism for describing structure within a function $g(X)$ defined over a set of variables $X = [x_1, \ldots, x_n]$ (most commonly a joint probability distribution or energy function over the $x_i$). Graphical models use this structure to organize computations involving $g(\cdot)$ and construct efficient algorithms for many inference tasks, including optimization to find a maximum a posteriori (MAP) configuration, marginalizing, or computing the likelihood of observed data. For the purposes of this paper, we assume that each variable $x_i$ takes on values from some finite set and focus primarily on the problem of marginalization.

### 2.1 Factor Graphs

Factor graphs [10] describe the factorization structure of the function $g(X)$ using a bipartite graph consisting of *factor* nodes and *variable* nodes. Specifically, suppose such a graph $G$ consists of factor nodes $F = \{f_1, \ldots, f_m\}$ and variable nodes $X = \{x_1, \ldots, x_n\}$, and let $X_j \subseteq X$ denote the neighbors of factor node $f_j$. Then, $G$ is said to be consistent with a function $g(\cdot)$ if and only if

$$g(x_1, \ldots, x_n) = \prod_j f_j(X_j).$$

In a common abuse of notation, we have used the same symbols to indicate both each variable node and its associated variable $x_i$, and similarly for each factor node and its associated function $f_j$.

It will often be convenient to refer to vertices without specifying whether they are variable or factor nodes. To this end, we define a set of artificial "factors" to be associated with both factors and variable nodes; for a generic vertex $v$ we define $\psi_v(X_v) \equiv 1$ for $v = x_i$, and $\psi_v(X_v) = f_j(X_j)$ for $v = f_j$.

### 2.2 Marginalization

A classic inference problem is that of marginalizing the function $g(X)$. Specifically, for some or all of the $x_i$, we are interested in computing the marginal function

$$g^i(x_i) = \sum_{X \setminus x_i} g(X).$$

When the factor graph representation of $g(X)$ is singly-connected (tree-structured), marginalization can be performed efficiently using sum-product [10]. In tree-structured graphs, sum-product is typically formulated as a two-pass sequence: rooting the tree at some node $v$, messages are sent upward (leaves to root), then back downward, after which one may compute the marginal for any node in the graph. In more general graphs (graphs with cycles), exact inference is less straightforward. One solution is to use a *junction tree* [11]; this first constructs a tree-structured hypergraph of $G$, then runs essentially the same inference process to compute marginals. The computational complexity of this process depends on the selected hypergraph and is exponential in the size of the cliques, or nodes of the hypergraph.

An alternate but essentially equivalent view of exact inference is given by the *bucket elimination* algorithm [5].

Bucket elimination chooses a sequence in which to marginalize the variables $x_i$, first multiplying together each of the factors which include $x_i$, then summing over $x_i$ to create a new factor and returning it to the pool. In tree-structured graphs, a marginal function $g^i(x_i)$ can be found in a manner similar to the upward pass of sum-product: rooting the tree at the node $x_i$ of interest, the summation operations are carried out first on the leaf nodes, followed by their parents, and so on until only the root $x_i$ remains. However, bucket elimination does not impose any particular elimination order, and we shall see in the sequel that alternative orders may come with other benefits.

Bucket elimination is closely related to junction tree based inference, and an equivalent junction tree may be defined implicitly by its specified elimination ordering [5].

### 2.3 RC-Trees for Adaptive Inference

In [2], an algorithm for adaptive inference in factor trees is described using "rake and compress" trees (RC-trees). The RC-tree data structure automatically selects an elimination ordering for the variables in the factor tree using a random-mate selection procedure, and stores functions at each node in the RC-tree representing sufficient statistics for its subtree. It was shown that construction of the RC-tree data structure requires time and space linear in the number of vertices $n$ of the factor graph, and produces a balanced tree with expected height $O(\log n)$.

The sufficient statistics stored in the RC-tree can be used to "query", or compute marginal distributions in the factor tree by passing information downward, taking at most expected $O(kd^{k+2} \log n)$ time, where $k$ is the maximal degree of the factor tree, and $d$ is the maximal dimension of each variable. Moreover, *changes* to the tree can also be incorporated in expected $O(kd^{k+2} \log n)$ time, including changes to the tree structure. The nature of the random-mate elimination ordering ensures that such changes affect only logarithmically many of the sufficient statistics.

Unfortunately, this formulation is restricted to tree-structured factor graphs, which limits its applicability in practice. In the following sections, we describe a generalization of the RC-tree structure which can cope with cycles in the factor graph while maintaining the desireable properties of the automatically chosen elimination ordering.

## 3 Hierarchical Clustering and Inference

We begin by describing a notion of hierarchical clustering in factor graphs which is compatible with but more general than that induced by RC-trees. We then describe how this clustering can be used to compute the marginal distribution at any vertex of the factor graph.

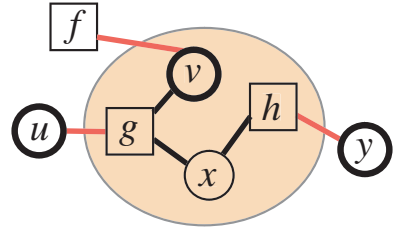

Figure 1: A cluster $C$ (shaded) with boundary edges (red) $\partial C = \{(u, g), (v, f), (y, h)\}$, boundary variables (bold circles) $X_c = \{u, v, y\}$ and cluster function $\varphi_C = \sum_{X \setminus X_c} g \cdot h = \sum_x g(u, v, x) \cdot h(x, y)$.

### 3.1 Hierarchical Clustering

For a factor graph $G = (X + F, E)$, a *cluster* $C$ is simply a set of vertices of $G$. We define the *boundary* of a cluster, written $\partial C$, as a set of edges with exactly one endpoint in $C$, and the *boundary variables* $X_C$ of $C$ to be the set of variables (variable nodes) incident to the boundary edges. For each cluster, we also define a *cluster function* $\varphi_C$ as the partial marginalization of all the factors in that cluster over all variables except the boundary variables:

$$\varphi_C(X_C) = \sum_{X \setminus X_C} \prod_{f_j \in C} f_j(X_j).$$

Fig. 1 shows an example cluster, its boundary and boundary variables.

We can then define a *hierarchical clustering* of $G$ to be a set of clusters $\mathcal{C} = C_1, \ldots, C_n$ such that the following conditions are satisfied:

1. Every vertex is covered by at least one cluster.

2. Clusters are nested: given two clusters either one is a subset of the other or they do not intersect. Moreover, if two clusters share a boundary edge, one is a subset of the other.

3. Each cluster $C$ has a unique identifier vertex $v$ : for any $C \in \mathcal{C}$ there is a unique $v \in C$ such that no other cluster contained by $C$ contains $v$. We write $\bar{v}$ to denote the cluster of identified with vertex $v$, i.e., $\bar{v} = C$.

4. For each maximal subcluster $C'$ of $C = \bar{v}$, i.e., $C'$ contained in no smaller cluster than $C$, there is an edge connecting $v$ and some $u \in C'$.

Fig. 2 shows a factor graph and a valid hierarchical clustering of the graph. Note that, by condition 3, the finest scale of the clustering are individual nodes.

A hierarchical clustering can be constructed bottom-up, by combining groups of sub-clusters which are adjacent to the same vertex. Since clusters are nested, we can represent a hierarchical clustering as a *cluster tree*, so that if a cluster $C'$ is a subset of $C$, then $C$ is an ancestor of $C'$ in the tree;

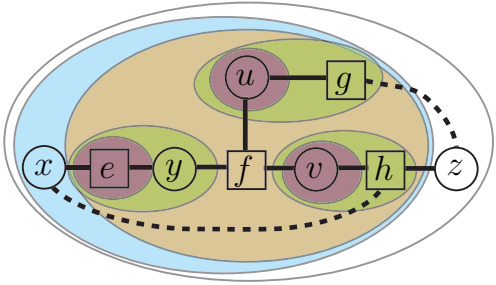

Figure 2: A factor graph $G$ and hierarchical clustering of $G$. Edges of $G$ designated as "non-tree" (see text) are shown as dashed.

the maximal subclusters of $C$ are the children of $C$. A cluster tree representation of the clustering in Fig. 2 is shown in Fig. 3. In the cluster-tree, each cluster is labeled based on its identifier vertex, e.g., the cluster $\bar{u}$ has identifier $u$. Also shown for each cluster are the boundary edges.

The cluster boundaries and their cluster functions can be computed in the cluster tree recursively, based on those of their immediate children. Let $S_{\bar{u}} = \{\bar{v}_1, \ldots, \bar{v}_k\}$ be the set of children of $\bar{u}$ in the cluster tree, and let $E(u)$ denote the edges containing $u$ as an endpoint. Then, the boundary of $\bar{u}$ is the set of edges that are in exactly one of $E(u), \partial\bar{v}_1, \ldots, \partial\bar{v}_k$, i.e.

$$\partial\bar{u} = E(u) \triangle \partial\bar{v}_1 \triangle \ldots \triangle \partial\bar{v}_k$$

where $\triangle$ is the symmetric set difference operator.

The cluster function for $\bar{u}$ can be computed as

$$\varphi_{\bar{u}}(X_{\bar{u}}) = \sum_{X \setminus X_{\bar{u}}} \psi_u(X_u) \prod_{\bar{v} \in S_{\bar{u}}} \varphi_{\bar{v}}(X_{\bar{v}}).$$

(Recall that the $\psi_u$ simply refer to factors of $g(\cdot)$.) Any such hierarchical clustering can be used to define a (partial) elimination ordering, with a variable being eliminated in the first (bottom-most) cluster which contains both the variable and all its neighboring factors. In the bucket elimination algorithm following this partial ordering, each cluster function $\varphi_C(X_C)$ then corresponds to the "new factor" created by marginalizing the factors in a given bucket.

Finally, we will find it useful to partition the edges of $G$ into two sets. In a hierarchical clustering $\mathcal{C}$, at each cluster $C = \bar{v}$ there exists at least one edge from $\bar{v}$ to each of its maximal subclusters $C'$ (if there is more than one, we can break ties arbitrarily). The collection of these edges form a subtree (or forest) of the original factor graph. We call these edges the "tree" edges $E_T \subseteq E$ of the hierarchical clustering; the remaining edges $E_N = E \setminus E_T$ we call the "non-tree" edges. In Fig. 2, the non-tree edges $E_N$ are shown as dashed.

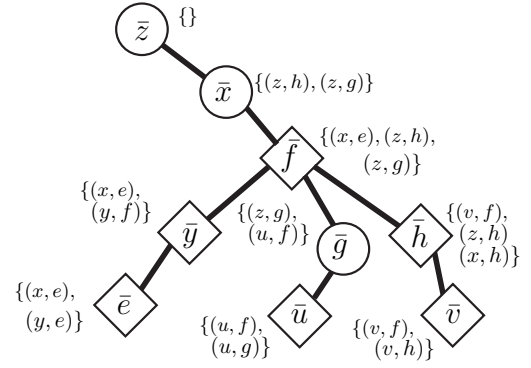

Figure 3: The cluster tree corresponding to Figure 2, showing the boundary of each cluster.

### 3.2 Computing Marginal Distributions

As with bucket elimination, the root of the cluster tree provides the marginal function for whatever variable is removed last. Moreover, it is also straightforward to compute the marginal at any other vertex by propagating information downward through the cluster tree. We compute the marginal distribution of a node $v$ as follows.

Let $\partial_T \bar{u}$ be the set of tree edges on the boundary of $\bar{u}$, i.e. $\partial_T \bar{u} = \partial\bar{u} \cap E_T$, and let $v_1, \ldots, v_n$ be the sequence from $\bar{v}$ to the root ($v_1 = \bar{v}$, $v_n$ the root). We compute a downward pass of marginalization functions from $v_n$ to $v_2$ as

$$M_{\bar{v}_i}(\cdot) = \sum_{X \setminus X_{\bar{v}_{i-1}}} \psi_{v_i}(\cdot) \prod_{\bar{u} \in A_{\bar{v}_i}} \varphi_{\bar{u}}(\cdot) \prod_{\bar{a} \in B_{\bar{v}_i}} M_{\bar{a}}(\cdot)$$

where $A_{\bar{v}_i} = S_{\bar{v}_i} \setminus \{\bar{v}_{i-1}\}$ is the set of children of $\bar{v}_i$ which are not on the path from $\bar{v}$ to the root, and $B_{\bar{v}_i}$ defined in terms of the tree edges as follows. If $\partial_T \bar{v}_i \setminus \partial_T \bar{v}_{i-1} = \{(a_1, a'_1), \ldots, (a_t, a'_t)\}$ with $a'_1, \ldots, a'_t \in \bar{v}_i$, then $B_{\bar{v}_i} = \{\bar{a}_1, \ldots, \bar{a}_t\}$. We know by the properties of the hierarchical clustering that each $\bar{a}_i \in B_{\bar{v}_i}$ is an ancestor of $\bar{v}_i$ in the cluster tree.

Each of these "messages" from parent $\bar{v}_i$ to child $\bar{v}_{i-1}$ is computed using only information on (messages into) the path above $\bar{v}_i$. The marginal at node $v$ is computed as

$$g^v(X_v) = \sum_{X \setminus X_v} \psi_v(\cdot) \prod_{\bar{u} \in S_{\bar{v}}} \varphi_{\bar{u}}(\cdot) \prod_{\bar{a} \in B_{\bar{v}}} M_{\bar{a}}(\cdot),$$

combining the information above and below $\bar{v}$.

In the previous work [2], the combination of $G$ being tree-structured and the selection criteria for creating clusters via rake or compress operations ensured that the computational complexity of each of these calculations was limited. For graphs with cycles, we shall see that these computations may grow more complex (due to the additional "non-tree" edges), but are still bounded and can be controlled sufficiently well to yield practically useful algorithms.

# 4 A Cluster Tree Data Structure

In this section, we describe a data structure for computing marginal distributions and performing various changes to the structure of the graphical model efficiently.

The idea behind our data structure is to maintain a balanced clustering of a factor graph. To do this, we require the user provide a factor graph along with a spanning tree (or forest) for that graph. We then build a hierarchical clustering of the factor graph, in which the specified spanning tree defines the tree edges $E_T$ of the clustering. Using this representation, we can perform marginal queries in time proportional to the depth of the cluster tree and to the size of the cluster functions stored at each node.

To compute and maintain a balanced clustering, we use the RC-Tree (Rake-and-Compress) tree data structure [1, 3]. This data structure constructs a hierarchical clustering of a tree by performing rake and compress operations and guarantees that the clustering has an expected depth of $O(\log n)$ in the size of the tree. The RC-Tree itself mimics the structure of the clustering: each node is a cluster and there is an edge from a cluster/node to its immediate subclusters. Thus, it enables traversing the clustering like an ordinary tree. In addition to these operations, RC-Trees enable inserting and deleting tree edges and updating the hierarchical clustering so that it remains balanced under any change to the underlying tree.

Since we work with general factor graphs, however, the RC-Tree representation itself does not suffice (RC-Trees are sufficient only for tree-structured factor graphs). To extend the representation, we follow the techniques described in Sec. 3 for computing the boundaries and cluster functions. More specifically, after building the clustering and its RC-Tree, we annotate each cluster with its set of boundary edges, including both tree and non-tree edges, and compute its cluster function as a partial marginalization of its factors over all variables except those on the boundary.

With an RC-tree annotated with boundaries and cluster functions, we can query the data structure to compute marginal functions in the manner described in Sec. 3.2.

To support changes to the underlying structure efficiently, we explicitly distinguish between tree edges and non-tree edges and we require that the spanning tree is kept consistent under changes. This requires, for example, that the user does not delete a spanning tree edge unless the graph becomes disconnected (i.e., there cannot be non-tree edges crossing the cut defined by that tree edge). In other words, the user is responsible for ensuring that the connectivity of the tree-edges matches the connectivity of the factor graph as a whole. This approach makes our interface somewhat crowded, but there is a reason: we wish to provide complete control to the user about the particular spanning tree being maintained, since this is crucial to performance (as we describe in Sec. 4.1). We note that distinguishing between tree and non-tree edges places no restrictions as to what changes can be performed, and the user can still insert and delete any edge. We simply require that if a tree edge is to be removed, it be replaced by another tree edge (perhaps by promoting a non-tree edge) unless its two endpoints are not connected via any other path. We handle changes to the structure of the factor graph as follows.

**Replacing a factor:** To replace a factor $f$, we first change it in the input factor graph. We then find the cluster $\bar{f}$ that identifies $f$ in the RC-Tree and update all cluster functions on the path from $\bar{f}$ to the root. Since each cluster function depends only on its subclusters, this sequence of updates suffices.

**Insert/delete non-tree edges:** Let $(u, f)$ be the non-tree edge being inserted or deleted. We first insert/delete $(u, f)$ into/from the input factor graph. We then find the clusters $\bar{u}$ and $\bar{f}$ in the RC-Tree and visit their ancestors in a bottom-up traversal. When visiting a cluster, we update its boundary edges, which may now need to be changed to exclude $(u, f)$ and recompute its cluster function based on its changed boundary. Since only ancestors of $\bar{u}$ and $\bar{v}$ may have $(u, v)$ as a boundary edge, updating only the ancestors suffices.

**Insert/Delete tree edges:** Let $(u, f)$ be the tree edge being inserted or deleted. We first insert/delete $(u, f)$ into/from the factor graph as requested. We then insert/delete $(u, f)$ from the spanning tree and use the change-propagation method supplied by the RC-Tree to update the clustering [1]. Change-propagation will update the RC-Tree by deleting some of the existing clusters and inserting some new clusters. We compute the boundaries and the cluster functions for newly created clusters by starting at the root(s) of the RC-Tree(s) involved in the operation and performing a top-down traversal until we visit all new clusters. It is a property of the RC-Tree data structure that all new clusters can be found in this way.

We note that it is for simplicity of presentation that we assume operations consisting of only single changes—multiple changes can be performed simultaneously.

## 4.1 Interface and Efficiency

We briefly describe the concrete interface to our data structure and analyze the running time for these operations.

The interface supports the following operations: ***cluster(G,T)***, ***query(v)***, ***replaceFactor(old, new)***, ***insertTreeEdge(e)***, ***deleteTreeEdge(e)***, ***insertTreeEdge(e)***, ***deleteNonTreeEdge(e)***. The ***cluster*** operation takes a factor graph $G$ and a spanning tree $T$ of $G$ and constructs hierarchical clustering. The ***query*** operation takes a vertex of the factor graph and returns the marginal of the vertex. The *replaceFactor* operation replaces a factor with another factor. The rest of the operations insert or delete edges in the input factor graph.

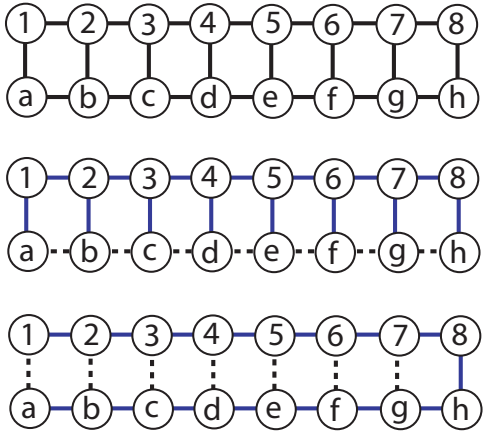

Figure 4: A pairwise factor graph (only variables shown) and two possible spanning trees (shown with thick edges). The first tree results in low measure $\mu_{T_1}(G) = 3$, but the second does not ($\mu_{T_2}(G) = 8$).

To analyze the efficiency of our data structure, we define a notion of the *measure* of a factor graph and its spanning tree. Let $G$ be a factor graph and $T$ a spanning tree; we first define the measure of an edge $e \in T$, written $\mu_T(e)$, as one plus the maximum size of the number of non-tree edges that cross a cut defined by $e$. More precisely, for an edge $e$ from $T$, let $T_e$ and $T'_e$ be the components of $T$ separated by deletion of $e$. Let $G_e$ and $G'_e$ be the subgraphs of $G$ induced by the vertices of $T_e$ and $T'_e$ respectively. Then $\mu_T(e)$ is the size of the cut between $G_e$ and $G'_e$. The measure of $G$ with respect to $T$, written $\mu_T(G)$, is the maximum-sized cut over all edges in $T$.

The importance of this measure is that it helps bound the size of the boundary for a cluster: if the number of tree edges that belong to the boundary of a cluster is $b$, then the boundary size is at most $b \cdot \mu_T(G)$. Since we use tree contraction to construct the cluster tree, our clusters have at most two tree edges in their boundary. Thus, the boundary of any of our clusters is at most $2\mu_T(G)$.

Fig. 4 shows a pairwise graphical model (top), with factors omitted (one for each edge), and two different spanning trees for it (middle and bottom) with spanning tree edges are highlighted. The factor graph has measure 3 with respect to the first spanning tree because removing any tree edge results in a cut of size at most 3. For example, for the edge $(4, d)$ the cut size is 3—it separates $d$ from the graph, which has two incident non-tree edges. Other vertical tree edges behave equivalently, and for the horizontal tree edges, the cut size is two. Thus for the first spanning tree the measure of the graph is small. For the second spanning tree, however, the measure is large. In particular, re-moving the edge $(8, h)$ separates the graph into two components consisting of the vertices at the top and those at the bottom with 8 cross edges. This example can be generalized to $n$ nodes such that the measure with respect to this kind of a spanning tree is $n/2$.

By allowing the user to choose the particular spanning tree being used, our data structure allows the measure of the graph to be kept small. This is important because as we prove in the next section, the measure the complexity of our data structure depends exponentially on $\beta$. In essence, these differences correspond to a good or poor choice of triangulation in the junction tree algorithm, or elimination orderings in bucket elimination. For these algorithms, good heuristics have been found by researchers over time, and are generally applied in an application-dependent manner.

For a factor graph $G$ and a spanning tree $T$, let $d$ be the domain of its variables and let $k$ the maximum degree of its nodes. We define the constant *characteristic* of $G$, denoted $\alpha$, as the constant $\alpha = d^{k+1}$. Note that representing an (input) factor itself may require this much space.

For the analysis consider some graphical model $G$ with spanning tree $T$, measure $\beta = \mu_T(G)$ and characteristic $\alpha$. Our bounds are in terms of the the characteristic and measure of $G$. For the bounds we assume that degree of the input graph $k$ and domain size of the variables $d$ are positive constants.

Our key lemma, stated below, bounds the time for computing the boundary and cluster function of a cluster.

**Lemma 4.1 (Cluster Cost)** *The boundary and cluster function of any cluster can be computed in $O(\alpha^\beta)$ time.*

**Proof:** We first note that since each cluster has at most two tree edges, it has a boundary of at most $2\beta$ edges.

Consider computing the boundary for some cluster. We will first bound the number of edges participating in the boundary computation. These edges consists of the boundary edges of the subclusters, the edges between the subclusters and the identifier vertex, and the boundary edges of the cluster itself. For counting purposes, suppose we place a pebble at each end point. The number of pebbles contributed by the $k$ subclusters is $2k\beta$. The number of pebbles contributed by the edges between the identifier and the subclusters is $k$, because the other endpoints of these edges are inside the clusters and already counted. Finally the pebbles contributed by the boundary edges of the cluster itself is $2\beta$ because one end point of the boundary edges is inside subclusters. The total number of edges is half the size of the pebbles, i.e., $\frac{2k\beta + 2\beta + k}{2} = (k+1)\beta + \frac{k}{2}$. By maintaining sorted boundaries and performing a $(k+1)$-way merge technique, we can compute the boundary for the cluster in $O\left(((k+1)\beta + \frac{k}{2})\log k\right)$ time. This running time is negligible compared to that of computing the cluster function, described next.

For computing the cluster function note that there can be at most $(k+1)\beta + \frac{k}{2}$ boundary variables, because each edge is incident on one variable. The combined domain of these variables then has size at most $d^{(k+1)\beta+\frac{k}{2}}$. We can compute the cluster functions by considering each member of the combined domain and performing $k$ additions or multiplications, giving total time $O(k \cdot d^{(k+1)\beta+\frac{k}{2}})$. ∎

**Theorem 1 (Hierarchical Clustering)** *Consider a factor graph $G$ with $n$ nodes and with spanning tree $T$. Let $\alpha$ be the characteristic of $G$ and let $\beta$ be the measure of $G$ with respect to $T$. We can compute the cluster tree of $G$ in $O(\alpha^\beta \cdot n)$ time The resulting cluster tree has $n$ clusters and expected $O(\log n)$ depth where the expectation is taken over internal randomization.*

**Proof:** It is known that the cluster tree can be computed in expected $O(n)$ time, independent of the cluster functions and boundaries [1, 3], and that the depth of the cluster tree is $O(\log n)$ in expectation. Since computing the boundary and the cluster function for each cluster takes $O(\alpha^\beta)$ time, the bound follows. ∎

We now state the theorem for queries and dynamic changes. Due to space restrictions, we omit the proofs here. Both theorems follow from the fact that changes and queries require traversing a path from the root to an update or a query node while perhaps updating cluster functions and boundaries or computing marginalization functions, which can be performed in $O(\alpha^\beta)$ time.

**Theorem 2 (Marginal Queries and Dynamic Changes)** *Consider a factor graph $G$ with $n$ nodes and with spanning tree $T$. Let $\alpha$ be the characteristic of $G$ and let $\beta$ be the measure of $G$ with respect to $T$. We can compute the the marginal of a variable in $O(\alpha^\beta \log n)$ expected time. Similarly each dynamic change can be processed in expected $O(\alpha^\beta \log n)$ time.*

## 5 Experimental Results

We compare the performance of a Matlab implementation of our algorithm to a standard implementation of a junction tree-based sum-product algorithm provided by the Bayes' Net Toolbox (BNT) [12]. We examine the speed-up provide by adaptive inference in two scenarios: synthetic data, which provides some control over the graph size and tree-width of the problems, and graphical models constructed from known protein backbone structures.

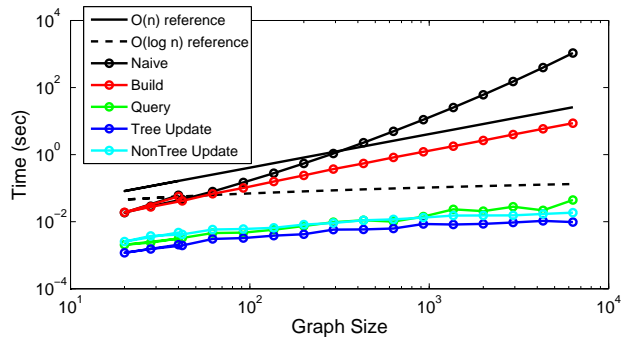

Figure 5: **Synthetic data.** Log-plot of the runtime of naive sum–product on a junction tree (BNT) versus our algorithm. Average update times (over 500 trials) are two to four orders of magnitude faster than performing inference from scratch.

### 5.1 Synthetic Data

For our synthetic data set, we randomly generated factor graphs with $n$ variables and $m$ factors, where $50 \leq n \leq 1000$, and $m = n - 1$. We initialize each input graph to be a simple Markov chain, where each factor $f_i$ depends on variables $x_i$ and $x_{i+1}$, where $1 \leq i < n$. This chain comprises the set of tree edges in our algorithm. Then, for given parameters $k$ and $\ell$, we add cycles by adding non-tree edges as follows: if $i$ is a multiple of $k$, we add variable $x_i$ to factor $f_{i+\ell-1}$ to create a cycle of length $\ell$. This creates a fairly structured yet loopy graph with limited tree-width.

The results for these synthetic experiments are shown in Fig. 5. We initially compute the (wall clock) time required to construct the cluster tree of the graph (using $k = 2$ and $\ell = 2$). To preserve the predictable tree-width of the problem, updates to the graph structure are performed in pairs by selecting a non-tree edge at random, removing it, updating the cluster tree, adding the edge back in and updating the cluster tree again. We also measure the time to query the marginal at a particular variable as well as the time to update factor definitions (i.e., the values of the factor and not the number of variables it depends on).

We find that our build time is slightly faster than direct inference using the BNT, possibly due to differences in elimination ordering, implicit (cluster tree) vs. explicit (junction tree) maintenance of the tree-decomposition, or simply differences in Matlab programming choices. Most importantly, we see that all of our update operations exhibit average running times (over 500 trials) that are logarithmic in $n$, and are between one to three orders of magnitude faster than performing inference from scratch.

### 5.2 Application to Protein Structure

Graphical models constructed from protein structures have been used to successfully predict structural properties [15]

| Protein | Size | BNT | Build | Query | Update | **Speedup** |
|---|---|---|---|---|---|---|
| 1aie | 31 | 0.213 | 0.165 | 0.008 | 0.012 | **8.24** |
| 1nkd | 59 | 0.422 | 0.252 | 0.011 | 0.012 | **18.0** |
| 1orc | 64 | 0.504 | 0.486 | 0.084 | 0.064 | **3.39** |
| 1vqb | 86 | 0.782 | 0.469 | 0.072 | 0.047 | **6.57** |
| 1rzl | 91 | 0.885 | 0.505 | 0.068 | 0.061 | **6.86** |

Figure 6: **Five proteins from the SCWRL benchmark.** Running times for an implementation of junction tree in BNT, and times for building, updating, and querying using our algorithm. Updates to factors and addition/removal of edges can be applied 3–18 times faster than recomputing from scratch.

as well as free energy [9]. These models are typically constructed by taking each node as an amino acid whose states represent *rotamers* [7], and basing conditional probabilities on a physical energy function (e.g., [14, 4]). A typical goal of using these models is to efficiently compute a maximum–likelihood (i.e. low–energy) conformation of the protein in its native environment. Updating factors allows us to study, for example, the effects of amino acid mutations, and the addition and removal of edges corresponds directly to allowing backbone motion in the protein. Furthermore, the effect of these updates on the model can then be incorporated in logarithmic time, which was not possible in previous approaches.

To test the feasibility of our algorithm for these applications, we constructed factor graphs from five moderately-sized proteins drawn from the SCWRL benchmark [4]. For each protein, we constructed the a factor graph by taking each amino acid as a variable, adding interactions between sequential amino acids as tree edges and steric interactions as non-tree edges. We performed the same updates as for our synthetic test set above. The table in Fig. 6 shows the results of our experiments. We see that for queries and updates, our approach gives a speedup of 3–18 times over inference from scratch. These results are consistent with our synthetic experiment above (i.e., graphs with just under a hundred variables), and show that an adaptive approach to inference can be useful in modeling protein structure. We note however, that for larger proteins, our choice of spanning tree (simply the protein backbone) produced graphs whose treewidth was too large for either our algorithm or sum-product. We are currently exploring other methods for choosing the spanning tree in a protein factor graph (e.g., based on rigid secondary structure elements).

## 6 Conclusion

We describe an efficient algorithm for adaptive inference in general graphical models. Our algorithm constructs a balanced representation of a spanning tree of the input graphical model, and represents cycles in the model by annotating this data structure. We can support all updates and marginal computations in expected $O(\alpha^\beta \log n)$ time, where $\alpha$ is a constant and $\beta$ is the size of a particular graph cut. Our experiments show that approach provides significant speedups on both synthetic and real protein data.